\documentclass{nature}

\usepackage{amsmath}
\usepackage{amssymb}
\usepackage{mathrsfs}
\usepackage{paralist}
\usepackage{hyperref}
\usepackage{booktabs}
\usepackage{graphicx}
\usepackage{float}
\usepackage{morefloats}
\usepackage{multirow}
\usepackage{color}
\usepackage{paralist}
\usepackage{todonotes}
\usepackage{wrapfig}
\usepackage{bbding}
\usepackage{tabularx}
\usepackage{enumitem}

\newcommand\blfootnote[1] 
{
  \begingroup
  
  \footnotetext{#1}
  \addtocounter{footnote}{-1}
  \endgroup
}

\title{Learning Multi-Site Harmonization of Magnetic Resonance Images Without Traveling Human Phantoms}

\author{Siyuan Liu\textsuperscript{1} and Pew-Thian Yap\textsuperscript{1,\Envelope}}
\begin{document}

\maketitle

\begin{affiliations}
 \item Department of Radiology and Biomedical Research Imaging Center (BRIC), University of North Carolina at Chapel Hill, NC, U.S.A.
\end{affiliations}

\begin{abstract}
Harmonization improves data consistency and is central to effective integration of diverse imaging data acquired across multiple sites.
Recent deep learning techniques for harmonization are predominantly supervised in nature and hence require imaging data of the same human subjects to be acquired at multiple sites. 
Data collection as such requires the human subjects to travel across sites and is hence challenging, costly, and impractical, more so when sufficient sample size is needed for reliable network training. Here we show how harmonization can be achieved with a deep neural network that does not rely on traveling human phantom data.
Our method disentangles site-specific appearance information and site-invariant anatomical information from images acquired at multiple sites and then employs the disentangled information to generate the image of each subject for any target site. We demonstrate with more than 6,000 multi-site T1- and T2-weighted images that our method is remarkably effective in generating images with realistic site-specific appearances without altering anatomical details. Our method allows retrospective harmonization of data in a wide range of existing modern large-scale imaging studies, conducted via different scanners and protocols, without additional data collection.
\end{abstract}

\blfootnote{\noindent \Envelope~Corresponding author: Pew-Thian Yap (\texttt{ptyap@med.unc.edu})}

\section*{Introduction}
Magnetic resonance imaging (MRI) is a non-invasive and versatile technique that provides good soft tissue contrasts useful for diagnosis, prognosis, and treatment monitoring.
Since MRI experiments are costly and time-consuming, modern large-scale MRI studies typically rely on multiple imaging sites to collaboratively collect data with greater sample sizes for more comprehensive coverage of factors that can affect study outcomes, such as age, gender, geography, socioeconomic status, and disease subtypes. Notable examples of multi-site studies include the Adolescent Brain Cognitive Development (ABCD)\cite{Jernigan2018The}, 
the Alzheimer's Disease Neuroimaging Initiative (ADNI)\cite{Mueller2005The}, and the Australian Imaging, Biomarkers and Lifestyle Flagship Study of Aging (AIBL)\cite{Ellis2009The}.

Multi-site data collection leads inevitably to undesirable elevation of non-biological variability introduced by differences in scanners\cite{Shinohara2017Volumetric} and imaging protocols\cite{Pomponio2020Harmonization}. 
Protocols can be prospectively harmonized by selecting for the individual sites the acquisition parameters that result in images with maximal inter-site consistency.
However, prospective harmonization requires extensive data acquisition for parameter tuning, needs to be performed before each study, and does not allow for correction of data collected in studies that have already taken place.
Moreover, significant inter-site variability can still occur in data collected with harmonized acquisition protocols simply due to irreconcilable differences between scanners\cite{Shinohara2017Volumetric}.
Retrospective MRI harmonization\cite{Yu2018Statistical} overcomes these limitations by performing post-acquisition correction and is hence applicable to existing studies for
improving the accuracy, reliability, and reproducibility of downstream processing, analyses, and inferences. 

\begin{table*}[!t]
	\renewcommand\arraystretch{1.5}
	\scriptsize
	\centering
	\caption{Existing MRI harmonization methods}
	\begin{tabularx}{\textwidth}{l|l|X|X}
		\specialrule{2pt}{8pt}{0pt}
		\multicolumn{2}{l|}{Category}                                                                                          & Method               & Summary        \\ \hline
		Statistics & Intensity normalization & Histogram matching\cite{He2013Intensity,Nyul1999On,Shah2011Evaluating} & Align distributions of voxel intensity values based on an image template constructed from several control subjects.\\ \cline{3-4} 
		& & White stripe\cite{Shinohara2014Statistical}  & Normalize intensity values based on patches of normal appearing white matter (NAWM). Rescaled intensity values are biologically interpretable as units of NAWM.\\ \cline{3-4}
		&  & Multi-site image harmonization by cumulative distribution function alignment (MICA)\cite{Wrobel2020Intensity} & Transform voxel intensity values of one scan to align with the intensity cumulative distribution function (CDF) of a second scan.\\ \cline{2-4}
		& Batch effect adjustment & Removal of Artificial Voxel Effect by Linear regression (RAVEL)\cite{Fortin2016Removing} & Estimate using a control region of the brain the latent factors of unwanted variation common to all voxels.\\ \cline{3-4} 
		&  & Combating batch effects (ComBat)\cite{Fortin2017Harmonization,Fortin2018Harmonization} & Identify batch-specific transformation to express all data in a common space. \\ \hline
		Learning  & Machine learning & Regression ensembles with patch learning for image contrast agreement (REPLICA)\cite{Jog2017Random} & Supervised training of a random forest for nonlinear regression of a target contrast.\\ \cline{3-4}
		& & NeuroHarmony\cite{Garcia-Dias2020Neuroharmony} & Supervised training of a random forest for nonlinear regression of a target contrast determined based on prescribed regions.\\ \cline{2-4}
		& Deep learning & DeepHarmony\cite{Dewey2019Deep} & Supervised training of a U-Net to produce images with a consistent contrast.\\ \cline{3-4} 
		& & Disentangled Latent Space (DLS)\cite{Dewey2020A} & Supervised training of a deep neural network using disentangled anatomical and contrast components.\\ \cline{3-4}
		& & Unlearning of dataset bias\cite{Dinsdale2021Deep} & Supervised training of a deep neural network to learn scanner-invariant features and then reducing scanner influence on network predictions in tasks of interest.\\ \cline{3-4}
		& & \textbf{MURD} & \textbf{Unsupervised} harmonization of images using content and style disentanglement learned from jointly multi-site images.\\ \cline{3-4}
		\specialrule{2pt}{-1.5pt}{-10pt}
	\end{tabularx}\label{tab:overview}
\end{table*}

\begin{figure*}[!t]
	\centering
	\includegraphics[width=.99\textwidth]{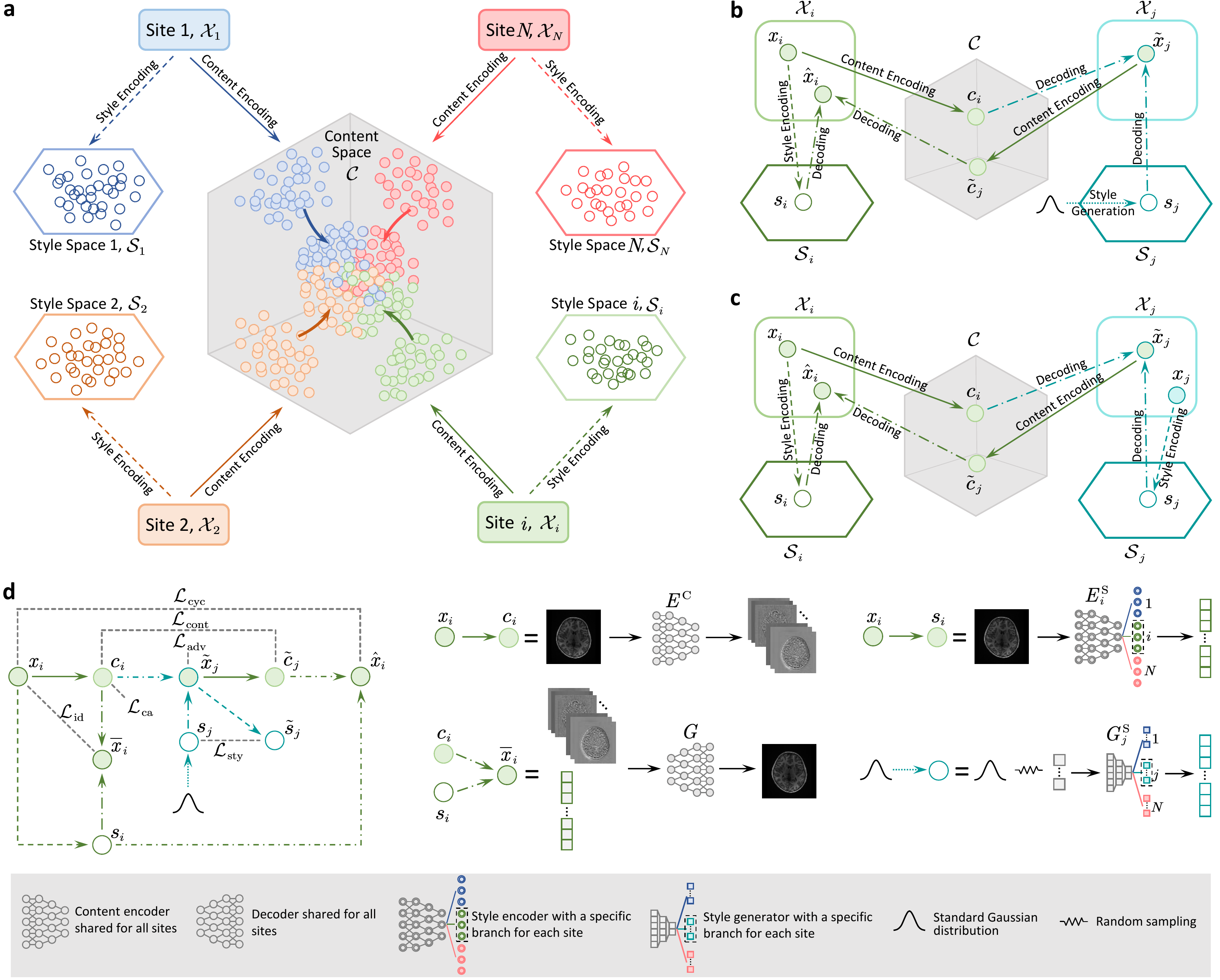}
		\vspace{-5pt}
	\caption{\textbf{Overview of MURD.}
		\textbf{a}, Multi-site representation disentanglement. Given image sets from $N$ sites $\{\mathcal{X}_i\}_{i=1}^N$, MURD disentangles each image set $\mathcal{X}_i$ in a site-invariant content space $\mathcal{C}$ and a site-specific style space $\mathcal{S}_i$ with the content features aligned in the content space $\mathcal{C}$. This is achieved using content-style disentangled cycle translation (CS-DCT).
		\textbf{b}, Site-specific CS-DCT. An image from the $i$-th site $x_i$ is encoded in content space $\mathcal{C}$ and style space $\mathcal{S}_i$ to obtain content features $c_i$ and style features $s_i$. Style features of the $j$-th site $s_j$ ($j\neq i$) are generated using style generator $G^{\text{S}}_j$. $s_j$ and $c_i$ decoded to generate a harmonized image for the $j$-th site $\tilde{x}_j$, which is in turn encoded to generate content features $\tilde{c}_j$ for reconstructing image $\hat{x}_i$ with style features $s_i$.
		\textbf{c}, Reference-specific CS-DCT. Unlike site-specific CS-DCT in \textbf{b}, style features $s_j$ are obtained by encoding a reference image $x_j$ in style space $\mathcal{S}_j$.
		\textbf{d}, The MURD framework. MURD implements CS-DCT by employing a site-shared content encoder, a site-specific style encoder, a site-shared generator, a site-specific style generator, and a site-specific discriminator for content encoding, style encoding, decoding, style generation, and adversarial learning, respectively. MURD is constrained by four types of losses: consistency losses $\mathcal{L}_\text{cyc}$, $\mathcal{L}_\text{cont}$ and $\mathcal{L}_\text{sty}$, adversarial loss $\mathcal{L}_\text{adv}$, content alignment loss $\mathcal{L}_\text{ca}$, and identity loss $\mathcal{L}_\text{id}$.
		\label{fig:framework}
	}
\end{figure*}

Existing retrospective harmonization methods are either 
statistics-based or learning-based (Table~\ref{tab:overview}).
%
Statistics-based methods align intensity distributions via intensity normalization\cite{Shah2011Evaluating,Nyul1999On,He2013Intensity,Shinohara2014Statistical,Wrobel2020Intensity} or batch effect adjustment\cite{Fortin2016Removing,Fortin2017Harmonization,Fortin2018Harmonization}.
However, these methods are typically limited to whole-image, but not detail-specific, harmonization.
Learning-based methods translate images between sites via nonlinear mappings determined using machine learning\cite{Jog2017Random,Garcia-Dias2020Neuroharmony} or deep learning\cite{Dewey2019Deep,Dewey2020A,Dinsdale2021Deep}, with or without supervision.
Machine learning methods predict harmonized images using regression models learned, typically with supervision,  with hand-crafted features.
In contrast, deep learning techniques automatically extract features pertinent to the harmonization task. Supervised methods typically require training data acquired from traveling human phantoms, which might not always be available for large-scale, multi-site, or longitudinal studies.
Unsupervised deep learning techniques\cite{Liu2017Unsupervised,Zhu2017Toward,Isola2017Image} determine mappings between sites using unpaired images and therefore avoid the need for traveling human phantom data. These methods are however unscalable to large-scale multi-site MRI harmonization as they typically learn pair-wise mappings between sites.
For $N$ sites, these methods learn $N(N-1)$ mappings for all site pairs and therefore require a large amount of data for learning the multitude of network parameters. These pair-wise methods are also ineffective by not fully and jointly utilizing complementary information available from all sites. 



Here, we draw inspiration from recent advancements in multi-domain image-to-image trans\-lation\cite{Anoosheh2018ComboGAN,Choi2018StarGAN,Choi2020StarGANv2} and introduce a unified  framework for simultaneous multi-site harmonization using only a single deep learning model. Our method, called multi-site unsupervised representation disentangler (MURD), decomposes an image into  anatomical content that is invariant across sites and appearance style (e.g., intensity and contrast) that is dependent on each site. Harmonized images are generated by combining the content of the original image with styles specific to the different sites.
More specifically, encoding an image in site-invariant and site-specific feature spaces is achieved via two encoders, i.e., a content encoder that captures anatomical structures common across sites and a style encoder that captures style information specific to each site.
A site-harmonized image is generated via a generator that combines the extracted content with the style associated with a target site. 
The target style can be specified by a reference image from a site or by a randomized style code generated by a style generator specific to each site. The latter allows multiple appearances to be generated in relation to natural style variations associated with each site.
The network is trained with losses that are designed to promote full representation disentanglement and to maximally preserve structural details. MURD is summarized in Figure~\ref{tab:overview} and detailed in the Methods section.

\section*{Results}
\begin{figure*}[!t]
	\centering
	\includegraphics[width=.94\textwidth]{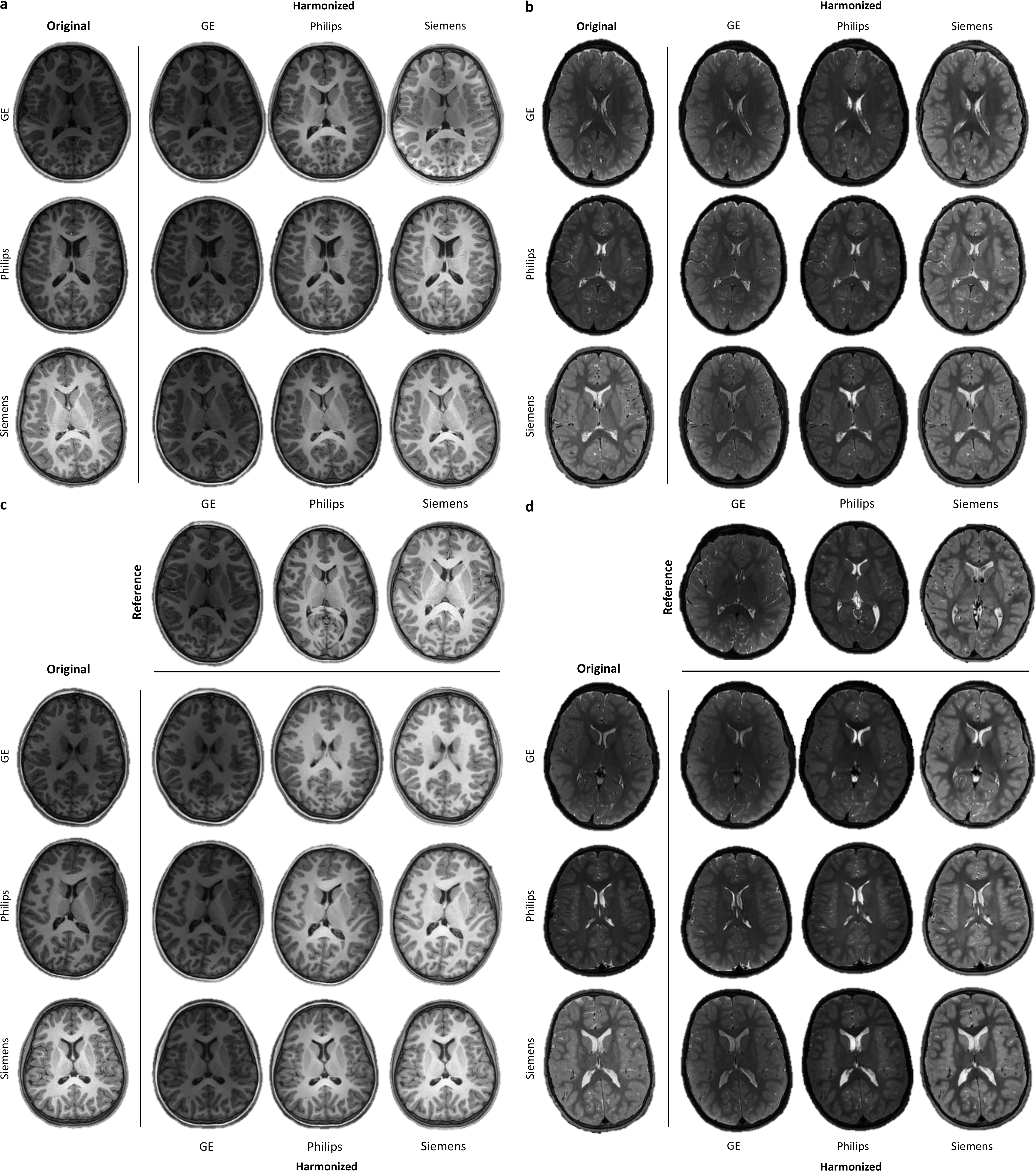}
	\caption{\textbf{Harmonization of T1- and T2-weighted Images.}
		Site-specific harmonization of \textbf{a}, T1-weighted images and \textbf{b}, T2-weighted images.
		Reference-specific harmonization of \textbf{c}, T1-weighted images and \textbf{d}, T2-weighted images.
		The original images are shown in the first column
		and the harmonized images are shown for GE, Philips and Siemens respectively in the second to fourth columns.
		MURD is remarkably effective in harmonizing contrasts and preserving details.\label{fig:gen_ref}}
\end{figure*}

\begin{figure*}[!t]
\centering
\includegraphics[width=0.95\textwidth]{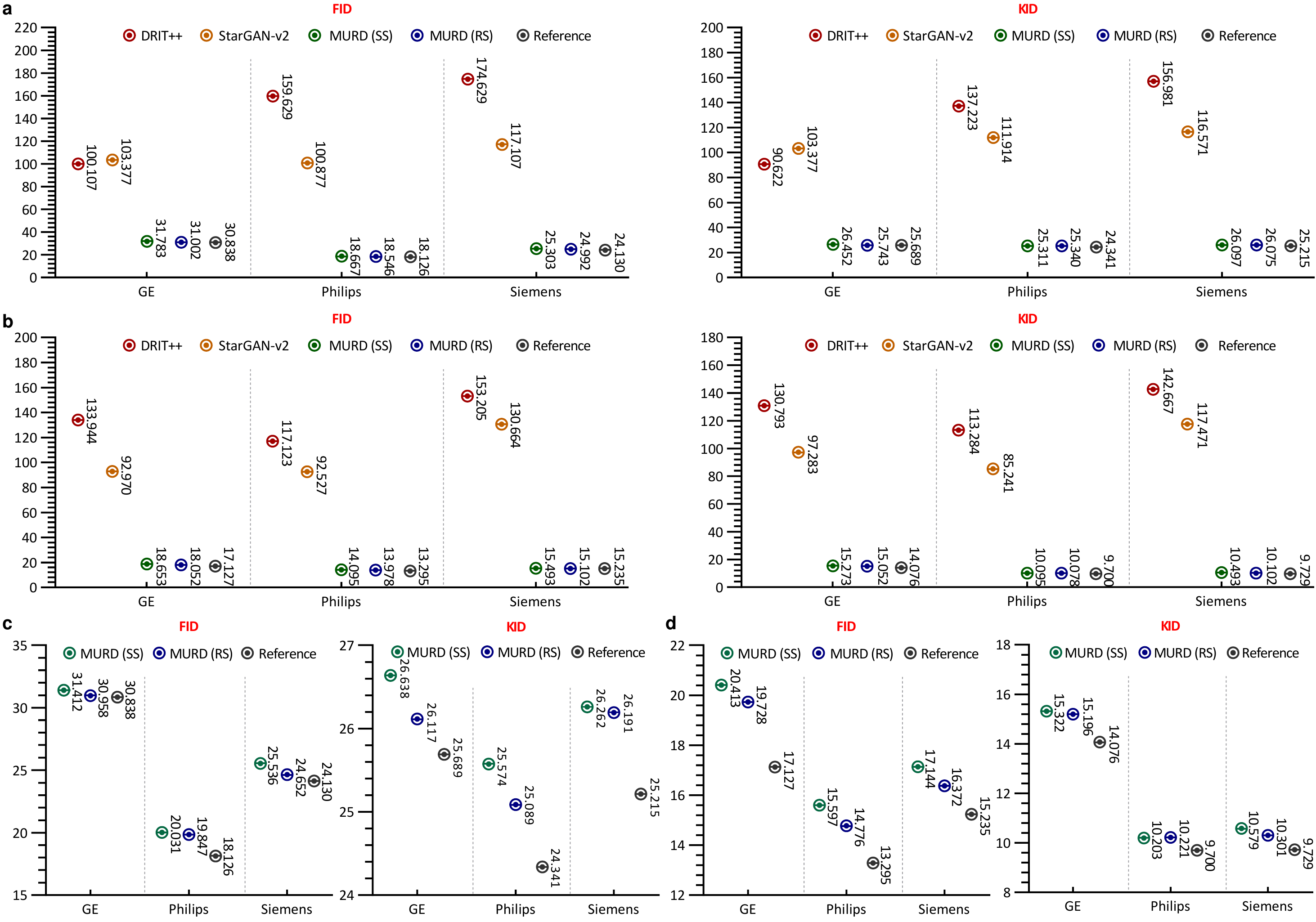}
\caption{\textbf{Numerical Evaluation of Harmonization Outcomes.}
	Quantitative evaluation conducted for the GE, Philips, and Siemens sites using FID and KID as metrics for \textbf{a}, T1-weighted images ($n=600$ slices per site) and \textbf{b}, T2-weighted images ($n=600$ slices per site) from the validation dataset, and \textbf{c}, T1-weighted images ($n=60000$ slices per site) and \textbf{d}, T2-weighted images ($n=60000$ slices per site) from the generalizability dataset. 
	Bullseyes show the means and error bars show the standard errors on the means.
	MURD, both site-specific (SS) and reference-specific (RS), yields lower FID and KID values that are closer to the reference values than DRIT++\cite{Lee2020DRIT} and StarGAN-v2\cite{Choi2020StarGANv2}. The FID and KID values for the generalizability dataset are largely consistent with those of the validation dataset.
\label{fig:comp_quan}
}
\end{figure*}

\begin{figure*}[!t]
\centering
\includegraphics[width=.95\textwidth]{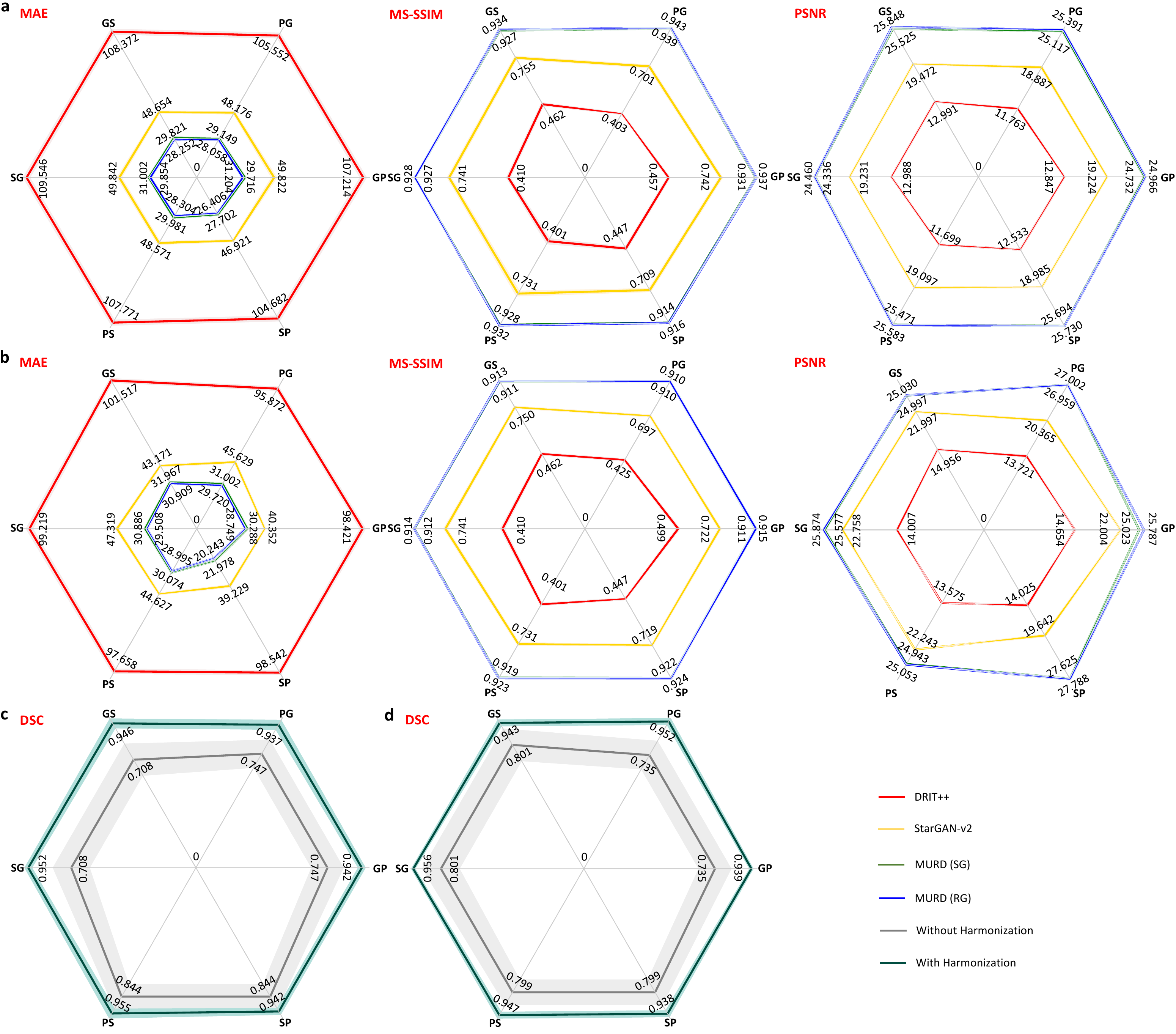}
\caption{
	\textbf{Validation via the Traveling Human Phantom Dataset.} 
	Evaluation of harmonized \textbf{a}, T1-weighted images and \textbf{b}, T2-weighted images from the traveling human phantom dataset for the following harmonization tasks: GE to Philips (GP, $n=60$ slices), Philips to GE (PG, $n=60$ slices), GE to Siemens (GS, $n=600$ slices), Siemens to GE (SG, $n=600$ slices), Philips to Siemens (PS, $n=120$ slices), and Siemens to Philips (SP, $n=120$ slices). 
	Segmentation consistency of \textbf{c}, T1-weighted images and \textbf{d}, T2-weighted images from the traveling human phantom dataset with and without harmonization: GP ($n=1$ volume), PG ($n=1$ volume), GS ($n=5$ volumes), SG ($n=5$ volumes), PS ($n=2$ volumes), and SP ($n=2$ volumes).
	The lines show the means and the shaded regions show the standard errors on the means. 
	MURD yields the best performance in terms of MAE, MS-SSIM, and PSNR. The improvement in DSCs indicates that MURD harmonization significantly increases consistency of tissue segmentation.
	\label{fig:ol_comp_quan}
}
\end{figure*}


\begin{figure*}[!t]
	\centering
	\includegraphics[width=.99\textwidth]{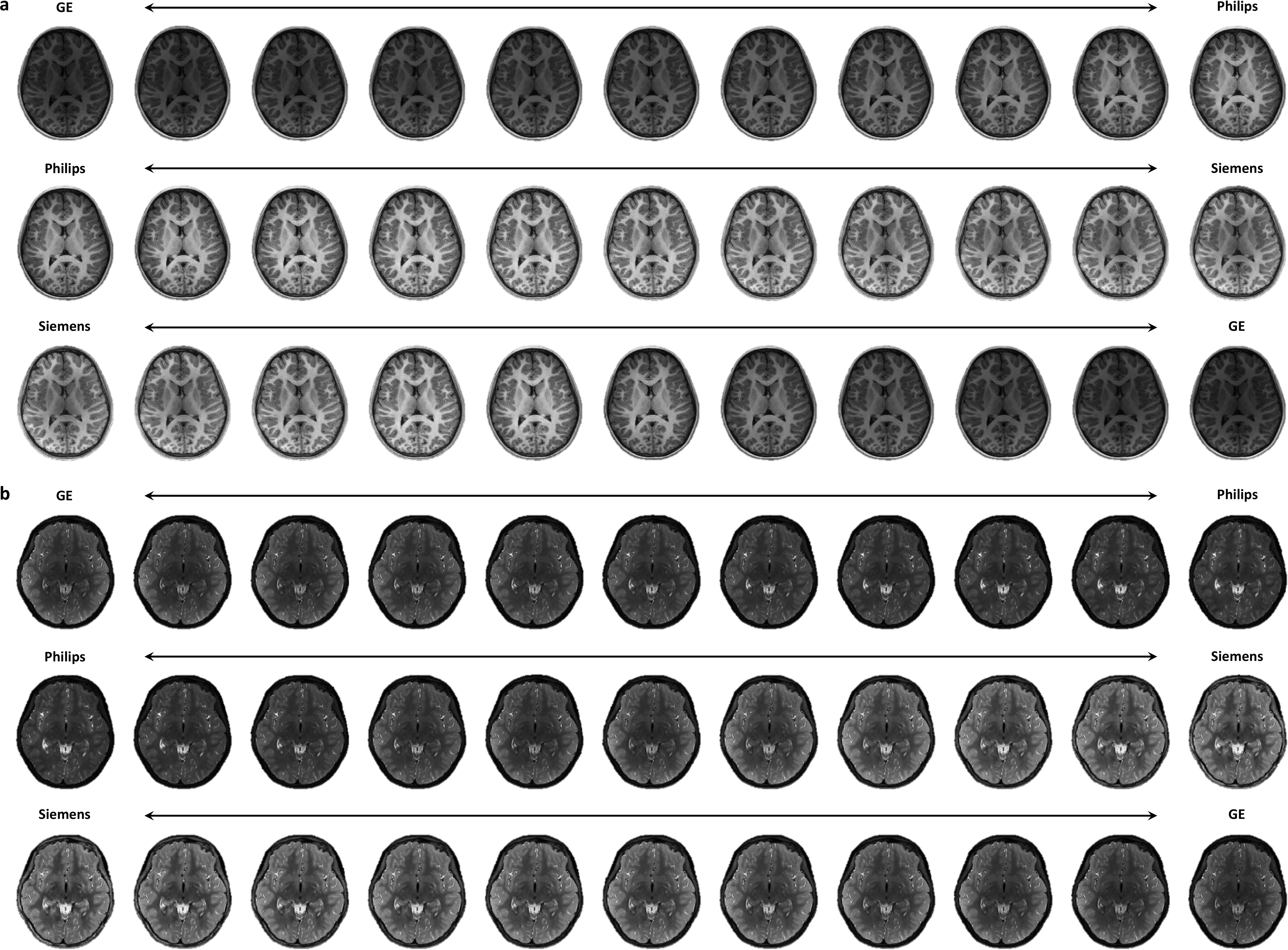}
	\caption{
		\textbf{Continuous and Incremental Harmonization.} 
		Continuous harmonization of \textbf{a}, T1-weighted images and \textbf{b}, T2-weighted images between two sites, demonstrating that 
		MURD smoothly changes image contrasts without altering anatomical details thanks to its ability to disentangle content and style information.\label{fig:continuous}}
\end{figure*}

\subsection{Materials}
We demonstrate the effectiveness of MURD on brain T1- and T2-weighted images of children 9 to 10 years of age
acquired through the Adolescent Brain Cognitive Development (ABCD) study\cite{Jernigan2018The} using scanners from different vendors, including General Electric (GE), Philips, and Siemens. Informed consents were obtained from all participants\cite{Jernigan2018The}.
Although the imaging protocols were matched across scanners at the different imaging sites\cite{Casey2018ABCD}, inter-scanner variability in appearance is still significant (Figure~\ref{fig:gen_ref}).
For simplicity, we group the images and treat them as coming from three \emph{virtual} sites---the GE, Philips, and Siemens sites.
We trained and tested MURD separately for T1-weighted images and T2-weighted images. For each modality, we trained MURD using a modest sample size of 20 images 
 per vendor. Three datasets per modality were used for testing different aspects of MURD:
\vspace{-\baselineskip}
\begin{enumerate}[itemsep=0pt,topsep=0pt,partopsep=0pt,parsep=0pt,label=(\roman{enumi})]
	\item \textbf{Validation Dataset} -- A small but diverse dataset of 10 images 
	per vendor, carefully selected to be structurally different from the images in the training dataset. The purpose of this testing dataset is to test the effectiveness of MURD beyond the training dataset.
	\item \textbf{Generalizability Dataset} -- A large dataset of 1000 images 
	randomly selected for each vendor. This testing dataset is allowed to be deliberately much larger than the training dataset to test the generalizability of MURD.
	\item \textbf{Human Phantom Dataset} -- A traveling human phantom dataset of 1 subject scanned on both GE and Philips scanners, 5 subjects scanned on both GE and Siemens scanners, and 2 subjects scanned on both Philips and Siemens scanners. This testing dataset allows numerical evaluation to ensure that anatomical structures are preserved after harmonization. Images for each subject were aligned using Advanced Normalization Tools (ANTS)\cite{Avants2009Advanced}.
\end{enumerate}
\vspace{-\baselineskip}
Note that the training dataset and the three testing datasets are mutually exclusive with no overlapping samples. 
Considering both modalities and all three vendors, more than 6000 images were used for evaluation. 60 central axial slices were extracted from each image volume for training and testing.

\subsection{MURD Harmonizes Contrasts and Preserves Details}
We demonstrate the effectiveness of MURD harmonization on
T1- and T2-weighted images (Figure~\ref{fig:gen_ref}). Note that MURD allows harmonization with respect to either a site
or a reference image. The former amounts to harmonization with respect to an image randomly drawn from a site image distribution. 
MURD is remarkably effective in adapting contrast and preserving details when harmonizing images from a source site with a target site (Figures~\ref{fig:gen_ref}a and b).
When the source and target sites are identical, MURD retain both contrast and details.
When given a reference image, MURD harmonizes the contrasts of images from a source site with the reference image but preserves the anatomical details of the original images (Figures~\ref{fig:gen_ref}c and d).


\subsection{MURD Outperforms State-of-the-Art Methods}
To further demonstrate the effectiveness and superiority of MURD, we compared it with two closely-related state-of-the-art unsupervised methods, i.e., DRIT++\cite{Lee2020DRIT} and StarGAN-v2\cite{Choi2020StarGANv2}, which are designed respectively for dual-domain and multi-domain image-to-image translation.
DRIT++ and StarGAN-v2 were implemented and trained as described in the original papers\cite{Lee2020DRIT,Choi2020StarGANv2}.
DRIT++ was trained once for every pair of sites. StarGAN-v2 was trained concurrently for multiple sites, similar to MURD.
We quantitatively compared the visual quality of the harmonized images using two metrics, i.e., the Frech\'{e}t inception distance (FID)\cite{Heusel2017Gans} and the kernel inception distance (KID)\cite{Binkowski2018Demystifying}, which reflect distribution discrepancy between two sets of images
in a manner that correlates well with human judgment\cite{salimans2016improved}. FID and KID are respectively computed based on the Frech\'{e}t distance and maximum mean discrepancy (MMD) of features from the last average pooling layer of the Inception-V3\cite{Szegedy2016Rethinking} trained on the ImageNet\cite{deng2009imagenet}. 
FID and KID were computed at the slice level for the harmonized images with respect to the training images of the target sites.
For site-specific harmonization, 10 target-site harmonized images were generated for each testing image of the source site with 10 randomly generated style codes.
For reference-specific harmonization, each image in the source site was harmonized with respect to 10 reference images randomly selected from the testing images of a target site.
%
The results (Figures~\ref{fig:comp_quan}a and b) indicate that MURD outperforms DRIT++ and StarGAN-v2 for both FID and KID. For comparison, reference FID and KID values were computed between the training and testing images from the same site (denoted as ``Reference"). 
The FID and KID values given by MURD are significantly closer to the reference values than DRIT++ and StarGAN-v2. 
MURD harmonization of the generalizability dataset (Figures~\ref{fig:comp_quan}c and d) yields FID and KID values that are highly consistent with the validation dataset and  close to the reference values. This indicates that, although trained using a modest dataset, the model is generalizable to a much larger dataset.


\subsection{MURD Efficacy Validated via Traveling Human Phantom Data}
The human phantom dataset allows direct quantitative evaluation of the effects of harmonization on consistency of structure and appearance.
Based on multiple metrics, including mean absolute error (MAE), multi-scale structural similarity index (MS-SSIM), and peak signal-to-noise ratio (PSNR), MURD significantly outperforms DRIT++ and StarGAN-v2, indicating better harmonization of contrast and preservation of anatomical details (Figures~\ref{fig:ol_comp_quan}a and b). 

\subsection{MURD Improves Tissue Segmentation Consistency}
Segmentation of brain tissues is sensitive to variation in image contrast and under- or over-segmentation might happen owing to differences in acquisition protocols.
We applied Brain Extraction Tool (BET)\cite{Smith2002Fast} and FMRIB's Automated Segmentation Tool (FAST)\cite{Zhang2001Segmentation} on T1- and T2-weighted images in the human phantom dataset for brain extraction and tissue segmentation.
Tissue segmentation consistency before and after harmonization was measured using the Dice similarity coefficient (DSC) with the tissue segmentation maps from the target site as references. The results (Figures~\ref{fig:ol_comp_quan}c and d) indicate that DSCs are improved remarkably by harmonization using MURD.

\subsection{MURD Supports Continuous and Incremental Harmonization}
We further demonstrate that MURD completely disentangles content and style information by visualization via continuous and incremental harmonization.
We generated images with between-site appearances to aid visual inspection of how appearance changes gradually between sites and whether unwanted anatomical alterations are introduced in the process.
Intermediate style features are calculated based on the style features of Site A, $s_\text{A}$, and Site B, $s_\text{B}$, via $(1-\beta)s_\text{A}+\beta s_\text{B}$, where $\beta\in [0,1]$.
The intermediate style features and the content features of an image are then used to generate an intermediate image. 
The results (Figures~\ref{fig:continuous}) indicate that MURD gradually and smoothly changes image appearance without altering anatomical details.

\subsection{Discussion}
We have demonstrated that MURD is remarkably effective in harmonizing MR images by removing non-biological site differences and at the same time preserving anatomical details. MURD network training involves only site-labeled images and requires no traveling human phantom data. This flexibility allows data from existing large-scale studies to be harmonized retrospectively without requiring additional data to be collected.

We have shown that MURD yields superior performance over DRIT++\cite{Lee2020DRIT} and StarGAN-v2\cite{Choi2020StarGANv2}.
For every pair of sites, DRIT++ embeds images in a site-invariant content space capturing information shared across sites and a site-specific style space. The encoded content features extracted from an image of one site are combined with style features from another site to synthesize the corresponding harmonized image. Learning is unsupervised and hence paired data is not required. However, DRIT++ is not scalable due to the need to learn all mappings for all site pairs.
DRIT++ is also ineffective because it cannot fully utilize the entire training data and can only learn from two sites at each time, causing it to miss global features that can be learned from images of all sites. 
Failure to fully utilize training data likely limits the quality of generated images. 
Unlike DRIT++, StarGAN-v2\cite{Choi2020StarGANv2} is scalable and performs image-to-image translations for multiple sites using only a single model. It has been applied to the problem of MRI harmonization\cite{Liu2021Style} with promising results.
In addition to greater scalability, StarGAN-v2 generates images of higher visual quality owing to its ability to jointly consider the information offered by images from all sites.  
StarGAN-v2, however, does not explicitly disentangle images into structural and appearance information. This introduces the possibility of altering anatomical details during harmonization via style transfer.
In contrast,  MURD enforces explicit disentanglement of  content and style features by jointly considering images from all sites, allowing it to produce harmonized images with diverse appearances with significantly better preservation of anatomical details (see Supplementary Figures~1--3).
Disentanglement safeguards harmonization against altering image anatomical contents and allows gradual and controllable harmonization via interpolation of style features. 


The harmonization target is specified for DRIT++ and StarGAN-v2 by a reference image. The appearance of the harmonized image is determined by the style features extracted from the reference image. 
In addition to a reference image, the harmonization target for MURD can be specified by a site label, which determines the output branch of the style generator and the style encoder. A latent code sampled from the standard Gaussian distribution determines an appearance specific to the site. 



A recent MRI harmonization method, called CALAMITI\cite{Zuo2021Information}, relies on intra-site supervised image-to-image translation and unsupervised domain adaptation for multi-site harmonization. 
This requires training a disentangled representation model with intra-site multi-contrast images (T1- and T2-weighted images) of the same subjects and retraining the model for a new site via domain adaptation\cite{Varsavsky2020Thomas}.
%
Unlike CALAMITI,  MURD requires only images from a single contrast and can learn multi-site harmonization simultaneously without needing fine-tuning or retraining. 




\begin{methods}
We consider the multi-site harmonization problem as image-to-image translation among multiple sites and propose an end-to-end multi-site unsupervised representation disentangler (MURD, see Figure~\ref{fig:framework}) to learn content-style disentangled cycle translation mappings that translate images forward and backward between any two sites. We employ two encoders to respectively embed each image in a site-invariant content space, which captures anatomical information, and a site-specific style space, which captures appearance information, and a generator to produce a harmonized image using the encoded content features and site-specific style features.

\subsection{Content-Style Disentangled Cycle Translation (CS-DCT)}
Given image sets $\{\mathcal{X}_i\}_{i=1}^N$ from $N$ distinct sites, MURD utilizes content-style disentangled cycle translation (CS-DCT) to disentangle each image set $\mathcal{X}_i$ in a site-invariant content space $\mathcal{C}$ and a site-specific style space $\mathcal{S}_i$ (Figure~\ref{fig:framework}a).
CS-DCT, realized with sequential forward and backward translation, can be site-specific (Figure~\ref{fig:framework}b) or reference-specific (Figure~\ref{fig:framework}c).
MURD jointly learns site-invariant content features $c_i\in\mathcal{C}$ and site-specific style features $s_i\in\mathcal{S}_i$ from image $x_i\in\mathcal{X}_i$, and utilize generator $G$ to construct the harmonized image $\tilde{x}_{j}\in\mathcal{X}_{j}$ ($j\neq i$) using content features $c_i$ and style features $s_{j}\in\mathcal{S}_j$ in forward translation.
Style features $s_j$ are generated by a style generator $G^{\text{S}}_j$ or extracted from a reference image $x_j\in\mathcal{X}_j$. In backward translation, MURD extracts the content features $\tilde{c}_{j}$ from the harmonized image $\tilde{x}_{j}$, which are in turn fed with style features $s_i$ to generator $G$ to reconstruct image $\hat{x}_i$, which is required to be consistent with the input image $x_i$.

\subsection{Multi-site Unsupervised Representation Disentangler (MURD)}
MURD implements CS-DCT in a end-to-end manner via five modules (Figure~\ref{fig:framework}d): 
\begin{enumerate}[itemsep=0pt,topsep=0pt,partopsep=0pt,parsep=0pt,label=(\roman{enumi})]
\item A content encoder $E^{\text{C}}$, shared for all sites, to extract content features from an image in a site-invariant space $\mathcal{C}$;
\item A style encoder $E^{\text{S}}_i$ for each site $i$ to extract style features from an image in a site-specific style space $\mathcal{S}_i$;
\item A generator $G$, shared for all sites, to synthesize images using content and style features;
\item A style generator $G^{\text{S}}_i$ for each site $i$ to yield style features $s_i$ that reflect the appearance style of images from the site;
and 
\item A discriminator $D_i$ for each site $i$ to distinguish between real and generated images.
\end{enumerate}
Specifically, given an input image $x_i$ from image set $\mathcal{X}_i$, content encoder $E^{\text{C}}$ and style encoder $E^{\text{S}}_i$, respectively, extract content features $c_i$ and style features $s_i$.
For a random site $j\neq i$, style generator $G^{\text{S}}_j$ takes a latent code $z$ randomly sampled from a standard normal distribution $\mathcal{N}(0,1)$ as input to create a $j$-site style features $s_j$.
With content features $c_i$ and style features $s_j$, generator $G$ constructs harmonized image $\tilde{x}_j$, which the discriminator $D_j$ then classifies as being either real or fake using an adversarial loss $\mathcal{L}_\textrm{adv}$:
\begin{equation}
\mathcal{L}_{_\textrm{adv}}=\mathbb{E}[\log D_i(x_i)]+\mathbb{E}[\log(1-D_j(\tilde{x}_j))].
\end{equation}
Content encoder $E^{\text{C}}$ and style encoder $E^{\text{S}}_j$ are used to extract content features $\tilde{c}_j$ and style features $\tilde{s}_j$ from $\tilde{x}_j$.
The consistency between $c_i$ and $\tilde{c}_j$ is enforced by a pixel-wise content consistency loss $\mathcal{L}_\text{cont}$:
\begin{equation}
\mathcal{L}_\text{cont}=\|c_i-\tilde{c}_j\|_1.
\end{equation}
The consistency between $s_j$ and $\tilde{s}_j$ is enforced by a pixel-wise style consistency loss $\mathcal{L}_\text{sty}$:
\begin{equation}
	\mathcal{L}_\text{sty}=\|s_i-\tilde{s}_j\|_1.
\end{equation}
Content site-invariance is enforced by a content alignment loss $\mathcal{L}_\text{ca}$:
\begin{equation}
\mathcal{L}_\text{ca}=\text{KL}(\mathcal{N}(c_i,I)\|\mathcal{N}(0,I)),
\end{equation}
where $\text{KL}(\cdot\|\cdot)$ is the Kullback–Leibler divergence and $I$ is an identity matrix.
Content features $c_i$ are randomly perturbed during the feed-forward step:
\begin{equation}
c_i=E^{\text{C}}(x_i)+\eta,\quad \eta\sim\mathcal{N}(0,I).
\end{equation}
The reconstructed image $\hat{x}_i$ of $x_i$ is produced by generator $G$ using content features $\tilde{c}_j$ and style features $s_i$. The consistency between $x_i$ and $\hat{x}_i$ is ensured by a pixel-wise and gradient-wise cycle consistency loss $\mathcal{L}_\text{cyc}$:
\begin{equation}
\mathcal{L}_\text{cyc}=\|x_i-\hat{x}_i\|_1+\lambda_\text{g}\|g(x_i)-g(\hat{x}_i)\|_1,
\end{equation}
where $g(\cdot)$ is the image gradient function and $\lambda_\text{g}$ is the loss weight for the gradient loss term.
Furthermore, an identity image $\bar{x}_i$ can also be constructed by generator $G$ using content features $c_i$ and style features $s_i$, which are identical to $x_i$ when $c_i$ and $s_i$ are completely disentangled.
We devise an identity loss to measure the pixel-wise difference between $x_i$ and $\bar{x}_i$ as
\begin{equation}
\mathcal{L}_\text{id}=\|x_i-\bar{x}_i\|_1.
\end{equation}
All modules are optimized with total loss function $\mathcal{L}$:
\begin{equation}\label{eq:TotalLoss}
\min_{E^{\text{C}},E^{\text{S}}_i,G,G^{\text{S}}_i}\max_{D_i} \mathcal{L}_\text{adv}+\lambda_\text{cont}\mathcal{L}_\text{cont}+\lambda_{ca}\mathcal{L}_\text{ca}+\lambda_\text{sty}\mathcal{L}_\text{sty}+\lambda_\text{cyc}\mathcal{L}_\text{cyc}+\lambda_\text{id}\mathcal{L}_\text{id},\quad i\in[1,N],
\end{equation}
where $\lambda_\text{cont}$, $\lambda_\text{ca}$, $\lambda_\text{sty}$, $\lambda_\text{cyc}$, and $\lambda_\text{id}$ are loss weights used for controlling the contributions of the respectively terms in the loss function.
Training ends when all modules are optimized, such that the optimized discriminator classifies the harmonized images into one of two categories with equal probability.
See Supplementary Figures~4--7 for the effects of the individual loss functions.

During inference, content features $c_i$ of input image $x_i$ are extracted using content encoder $E^{\text{C}}$, and style features $s_j$ are either generated using style generator $G^{\text{S}}_j$ or extracted from a reference image $x_j$. Harmonized image $\tilde{x}_j$ is created by generator $G$ using $c_i$ and $s_j$.

\subsection{Network Architecture}
The architectures of the components of MURD are 
described next.

\textbf{Content Encoder} Content encoder $E^{\text{C}}$ is shared among all sites and extracts the content features $c_i$ of an input image $x_i$ through three convolutional blocks and 4 residual blocks.
Each convolutional block is composed of three sequential layers, i.e., convolution, instance normalization (IN)\cite{Ulyanov2017Improved}, and leaky ReLU (LReLU) activation.
Each residual block consists of a nonlinear mapping branch and a shortcut branch. The nonlinear mapping branch is constructed by a series of layers, i.e., convolution, IN, LReLU, convolution, and IN. The shortcut branch is an identity mapping of the block input.
We use an IN layer instead of a batch normalization layer\cite{Ulyanov2017Improved} to accelerate model convergence and maintain independence between features.
All normalized feature maps are activated by LReLU with a negative slope of 0.2.

\textbf{Style Encoder} Style encoder $E_i^{\text{S}}$ is composed of site-shared and site-specific subnetworks.
The site-shared subnetwork is constructed by a convolution layer, four pre-activation residual blocks, and a global average pooling layer.
The site-specific subnetwork is composed of $N$ fully connected layers corresponding to the $N$ individual sites.
The pre-activation residual block is constructed by integrating LReLU activation followed by a convolution layer with unit stride into a residual block, where an average pooling layer is adopted to downsample the intermediate features and the shortcut branch is implemented by an average pooling layer and a convolution layer with unit kernel size and stride. 
Note that we extract style features without IN layers since IN removes feature means and variances, which contain important style information.
The output dimension is set to 64. Style features $s_i$ have the same dimension.

\textbf{Generator} Site-shared generator $G$ merges content features $c_i$ and style features $s_j$ to create a harmonized image $\tilde{x}_j$ using four residual blocks identical to the content encoder, two upsampling blocks and a convolution layer with hyperbolic tangent (tanh) activation. The upsampling block consists of deconvolution, IN, and LReLU activation.

\textbf{Style Generator} Style generator $G^{\text{S}}_j$ consists of a multilayer perception (MLP) with $N$ output branches.
Six fully connected layers are shared among all sites, followed by a fully connected layer for each site.
We set the dimensions of the latent code, the hidden layer, and the style features to 16, 256, and 64, respectively.
We randomly sample the latent code $z$ from the standard Gaussian distribution.

\textbf{Discriminator} Discriminator $D_j$ consists of site-shared and site-specific subnetworks, similar to the style encoder. 
Specifically, three convolutional blocks and a global average pooling are shared among all sites, followed by a specific fully connected layer for real/fake classification for each site. 

\subsection{Implementation Details} 
MURD was implemented using Tensorflow. Evaluation was based on a machine with a CPU (i.e., Intel i7-8700K) and a GPU (i.e., NVIDIA GeForce GTX 1080Ti 11GB). The ADAM optimizer with $1\times10^{-4}$ learning rate was utilized for minimization based on the loss function. For all datasets, we used $\lambda_\text{cont}=10$, $\lambda_\text{ca}=0.01$, $\lambda_\text{sty}=10$, $\lambda_\text{cyc}=10$, $\lambda_\text{g}=0.1$, and $\lambda_\text{id}=10$ for the corresponding losses.
For all datasets, every three adjacent slices in each volume were inserted into three channels. Each channel
was then normalized to have a range between -1 and 1 and zero-padded to 256$\times$256.


\end{methods}

\begin{addendum}
	\item[Acknowledgments] This work was supported in part by the United States National Institutes of Health (NIH) grants EB006733 and MH125479. 
	\item[Competing Interests] The authors declare no competing interests.
	\item[Correspondence] Correspondence and requests for materials
	should be addressed to Pew-Thian Yap~(email: ptyap@med.unc.edu).
	\item[Data Availability] The Adolescent Brain Cognitive Development (ABCD) 2.0 data used in this study can be obtained from the National Institutes of Mental Health Data Archive (NDA) (\url{http://nda.nih.gov/}).
	\item[Author Contribution] S.L. designed the framework and network architecture, carried out the implementation, performed the experiments, and analyzed the data. S.L. and P.-T.Y. wrote and revised the manuscript.
	P.-T.Y. conceived the study and were in charge of overall direction and planning. 
\end{addendum}

\subsection{References}
\footnotesize

\end{document}